\documentstyle[preprint,eqsecnum,aps,twoside,psfig,epsf]{revtex}

\renewcommand\vector[1]{\mbox{\boldmath{$#1$}}}

\newcommand\per[1]{\vector{#1}_{\perp}}
\newcommand\nhat{\hat{\vector{n}}}
\newcommand\ndot[1]{\vector{#1}\cdot\nhat}
\newcommand\energy[1]{\sqrt{\vector{#1}^{2} + m^{2}}}
\newcommand\measure{\left[\int\right]}

\newcommand\fig[1]{Fig.~\ref{#1}}
\newcommand\tab[1]{Tab.~\ref{#1}}
\newcommand\SEC[1]{Section~\ref{#1}}
\newcommand\eqn[1]{Eq.~(\ref{#1})}

\newcommand\average[1]{{\rm Re}\left\{\tilde{M}_{#1}^{(1)}\right\}}
\newcommand\swave[1]{\average{#1,S}}

\renewcommand{\theequation}{\thesection-\arabic{equation}}

\begin{document}

\draft
\preprint{SNUTP-98-024}
\title{The Rotation Average in Lightcone Time-Ordered Perturbation Theory}
\author{
   Chueng-Ryong Ji \\
      {\it Department of Physics, North Carolina State University} \\
      {\it Raleigh, North Carolina 27695-8202} \\
   Gwang-Ho Kim and Dong-Pil Min \\
      {\it Department of Physics, Seoul National University} \\
      {\it and Center for Theoretical Physics} \\
      {\it Seoul 151-742, Korea}
}
\date{\today}
\maketitle

\begin{abstract}
We present a rotation average of the two-body scattering amplitude in the
lightcone time($\tau$)-ordered perturbation theory. Using a rotation average
procedure, we show that the contribution of individual time-ordered
diagram can be quantified in a Lorentz invariant way.
The number of time-ordered diagrams can also be reduced by half
if the masses of two bodies are same.
In the numerical example of $\phi^{3}$ theory, we find that the higher 
Fock-state contribution is quite small in the lightcone quantization. 
\end{abstract}

\section{Introduction}\label{sec:SEC_intro}

The invariant amplitude obtained by calculating a covariant
Feynman diagram can equivalently be given by the sum of 
the corresponding time-ordered diagrams in the old fashioned
perturbation theory(OFPT). As it is well known\cite{ref:Brodsky,ref:Weinberg},
the individual time-ordered 
diagram is not invariant under some of the Lorentz transformations,
{\it e.g.}, boost or rotation,
while the covariant Feynman diagram is completely Lorentz invariant.
Under which 
part of the Lorentz transformations the individual diagram is not invariant 
depends on whether we use
the ordinary equal-$t$ quantization or 
the lightcone equal-$\tau$ quantization where $\tau = t+z/c$.  
The Poincar\'e algebra in these two 
schemes are significantly different. It is often remarked that
in the equal-$t$ quantization the 
boost operation is dynamic and the rotation is kinematic,
while in the equal-$\tau$ quantization the rotation is 
dynamic and the boost operation is kinematic\cite{ref:Leutwyler}.
These significantly different features of Poincar\'e algebra in two
schemes lead to the non-invariance of the individual diagram  under
different part of Lorentz transformations. 
In the equal-$t$ quantization, the individual diagram is not invariant under 
the boost 
transformation, while in the equal-$\tau$ quantization the individual diagram 
is not invariant under the rotation. However, it is crucial to note that
the property of rotation is very 
different from the property of boost operation because the rotation is 
compact {\it i.e.}, closed and periodic,
while the boost operation is open and not periodic. Thus, one may take 
advantage of 
the rotation in equal-$\tau$ quantization. 
Already, M. Fuda\cite{ref:fuda1} suggested the angular averaging of the 
potential as a way of restoring Poincar\'e invariance in the explicit
example of $\pi N$ scattering problem. We have also realized that the
physical on-shell partial wave amplitudes presented in Ref.\cite{ref:fuda2}
were in fact identical to the rotation average of the lightcone scattering
partial wave amplitudes\cite{ref:JKM}.
In this paper, we give an example of rotation advantage in OFPT. 
If we make a rotation average of the 
individual diagram, then the result is of course invariant 
under rotation and thus
the individual diagram can be made invariant under the rotation.
The similar average procedure for the boost operation cannot be 
made in the equal-$t$ quantization because the parameter space of 
boost, {\it i.e.}, velocity is not compact.  As we will show explicitly in 
this work, the individual $\tau$-ordered diagram can be made invariant 
under the entire Lorentz transformation using an average procedure.
Furthermore,
in the calculations of the two-body scattering amplitudes
where the masses of two bodies are same,
one does not need to calculate the 
entire number of $\tau$-ordered diagrams but to calculate only the 
half of the entire number of diagrams because the half of total number of 
diagrams is reproduced by the other half. Thus, one can evaluate 
the magnitude of each diagram in the Lorentz invariant way
once the average procedure is fixed. 

In the example of this work, we 
found that the higher Fock-state contribution is very small in the 
lightcone quantization. Our nontrivial point is that this smallness can 
be asserted in a reference-frame independent way. 
Without loss of generality but for simplicity, we 
show this point using an explicit example of Feynman amplitudes
in $\phi^3$ theory\cite{ref:Wick}.
However, our method is
generic to the equal-$\tau$ quantization scheme, and thus applicable to
any other field theory.
In this work, we calculate the lowest order 
two-body interaction diagrams shown in Figs.~\ref{fig:0,CVPT}--\ref{fig:0,LT_OFPT} and the real part of one 
higher order ladder diagram shown in Figs.~\ref{fig:1,CVPT} and \ref{fig:1,LT_OFPT}.
The generation of these diagrams from the covariant Bethe-Salpeter kernel
was discussed in Ref.\cite{ref:BJS}. A general algorithm of producing
the $\tau$-ordered diagrams from any Feynman diagram was also recently
presented by Ligterink and Bakker\cite{ref:LB}.
In \SEC{sec:SEC_anal...}, we present analytic calculations of Feynman diagrams shown 
in Figs.~\ref{fig:0,LT_OFPT}~and~\ref{fig:1,LT_OFPT}. In \SEC{sec:SEC_num...}, the numerical computations are made and 
the results are summarized. Conclusions and discussions follow in \SEC{sec:SEC_con...}.
In the Appendix, the equivalence is shown between the covariant next-to-leading order ladder
diagram and the sum of $\tau$-ordered diagrams\cite{ref:BJS,ref:LB}.

\section{Scattering amplitude}\label{sec:SEC_anal...}

In the $\phi^{3}$ covariant perturbation theory (CVPT), the lowest order
Feynman amplitude for the two-body scattering is given by the single
diagram shown in \fig{fig:0,CVPT}. This single diagram in \fig{fig:0,CVPT}
corresponds to the sum of two diagrams shown in \fig{fig:0,ET_OFPT} in
the ordinary time($t$)-ordered OFPT.
However, as we have discussed in the Introduction (\SEC{sec:SEC_intro}), each
separate diagram in \fig{fig:0,ET_OFPT} is not boost invariant even though
it is invariant under rotation. Only the sum of the two diagrams is
completely Lorentz invariant. Now, let's consider changing the time in
OFPT from $t$ to the lightcone time $\tau=t+\ndot{x}/c$ where $\nhat$
is a unit vector on the lightcone surface ({\it e.g.}, $\tau=t+z/c$
means $\nhat=\hat{z}$). If we change from $t$ to $\tau$, then
we still have two $\tau$-ordered diagrams as shown in \fig{fig:0,LT_OFPT}
which apparently look identical to those in \fig{fig:0,ET_OFPT}.
However, each diagram in \fig{fig:0,LT_OFPT} depends on $\nhat$ and
one can easily find that it is not invariant under rotation but nevertheless
invariant under boost. This drastic change of the Lorentz property from
the case of \fig{fig:0,ET_OFPT} is exactly what allows us to make an average
of each diagram. By taking advantage of compactness in the rotation, we
now take the average value of each diagram in \fig{fig:0,LT_OFPT} over $\nhat$.

After diagrams are averaged over $\nhat$, the restoration of the rotational
symmetry is manifest for each diagram. Of course, the sum
of diagrams remains same whether we take the average over $\nhat$ or not.
If the two particles of mass $m$ scatter with the initial(final) c.m.
momentum $\vector{k}(\vector{l})$, then the scattering amplitudes,
$M_{i}^{(0)}(\vector{k},\vector{l},\nhat), i=1,2$ for the two diagrams
in \fig{fig:0,LT_OFPT},
are given by (modulo a common constant factor);
\begin{eqnarray}
M_{1}^{(0)}(\vector{k},\vector{l},\nhat)
    & = & F_{1}(x,\per{k};y,\per{l}) \ , \\
M_{2}^{(0)}(\vector{k},\vector{l},\nhat)
    & = & F_{1}(1-x,-\per{k};1-y,-\per{l}) \ , \\
    & = & F_{1}(1-x,\per{k};1-y,\per{l}) \ ,
\end{eqnarray}
where
\begin{equation}
F_{1}(x,\per{k};y,\per{l}) = \frac{\theta(x-y)}{x-y}
\left[ \frac{\per{k}^{2}+m^{2}}{x} - \frac{\per{l}^{2}+m^{2}}{y}
     - \frac{(\per{k}-\per{l})^{2} + \mu^{2}}{x-y} \right]^{-1} \label{eq:F1}
\end{equation}
with $\mu$ being the mass of exchanged particle, and
\begin{eqnarray}
x       & \equiv & \frac{k^{+}}{P^{+}}
          = \frac{1}{2} \left[ 1 + \frac{\ndot{k}}{\energy{k}} \right] \ ,
            \label{eq:x_frac} \\
y       & \equiv & \frac{l^{+}}{P^{+}}
          = \frac{1}{2} \left[ 1 + \frac{\ndot{l}}{\energy{l}} \right] \ , \\
\per{k} & \equiv & \vector{k} - (\ndot{k})\nhat \ , \\
\per{l} & \equiv & \vector{l} - (\ndot{l})\nhat \ . \label{eq:l_per}
\end{eqnarray}
From Eqs. (\ref{eq:x_frac}) - (\ref{eq:l_per}), one can easily note that
$(x;\per{k};y;\per{l}) \longrightarrow (1-x;\per{k};1-y;\per{l})$ as
$\nhat \longrightarrow -\nhat$.
Also, $F_{1}$ depends only on the relative sign of $\per{k}$ and $\per{l}$.
Thus, if we take the average of $M_{i}^{(0)}(\vector{k},\vector{l},\nhat)$
over $\nhat$ and define $\tilde{M}_{i}^{(0)}(\vector{k},\vector{l})$ as
\begin{equation}
\tilde{M}_{i}^{(0)}(\vector{k},\vector{l}) \equiv
\frac{1}{4\pi} \int d\nhat M_{i}^{(0)}(\vector{k},\vector{l},\nhat) \ ,
\end{equation}
then we find
\begin{equation}
\tilde{M}_{1}^{(0)}(\vector{k},\vector{l}) =
\tilde{M}_{2}^{(0)}(\vector{k},\vector{l}) \ ,
\end{equation}
because
\begin{equation}
M_{2}^{(0)}(\vector{k},\vector{l},-\nhat) =
M_{1}^{(0)}(\vector{k},\vector{l},\nhat) \ .
\end{equation}
We may summarize out results for the lowest order as follows:
\begin{equation}
M_{\rm SUM}^{(0)}(\vector{k},\vector{l})
 = \sum_{i=1}^{2} M_{i}^{(0)}(\vector{k},\vector{l},\nhat)
 = \sum_{i=1}^{2} \tilde{M}_{i}^{(0)}(\vector{k},\vector{l})
 = 2 \tilde{M}_{1}^{(0)}(\vector{k},\vector{l}) \ .
\end{equation}

From this, we notice that,
after averaging over $\nhat$, we not only restore the rotional symmetry
for each separate diagram in \fig{fig:0,LT_OFPT} but also we can actually
reduce the number of diagrams necessary for the calculation by half for the two-body scattering
amplitude. The reason for the reduction in the number of diagrams is due to the fact that
$\nhat \longrightarrow -\nhat$ corresponds to
$x \longrightarrow (1-x)$ and $y \longrightarrow (1-y)$ and the two-body scattering
amplitude must be symmetric under this change of variables.
In order to show an explicit example beyond the leading order, let's now
consider the next-to-leading order ladder diagram in CVPT as shown in \fig{fig:1,CVPT}.

While in the $t$-ordered OFPT there are $4!=24$ diagrams,
we have only 6 diagrams in the $\tau$-ordered OFPT
(See \fig{fig:1,LT_OFPT}.). For example, a diagram shown in \fig{fig:1,NULL}
appears in the $t$-ordered OFPT but not in the $\tau$-ordered OFPT\cite{ref:Weinberg,ref:Chang}.

In this next-to-leading order, the scattering amplitudes
$M_{i}^{(1)}(\vector{k},\vector{l},\nhat),i=1,2,\cdots,6$
for the six diagrams in \fig{fig:1,LT_OFPT} are given
in the $\tau$-ordered OFPT by
\begin{eqnarray}
M_{1}^{(1)}(\vector{k},\vector{l},\nhat) & = & \measure
    \frac{F_{1}(x,\per{k};z,\per{q})F_{1}(y,\per{l};z,\per{q})}
         {F_{0}(x,\per{k};z,\per{q})} \ , \label{eq:M_1} \\
M_{2}^{(1)}(\vector{k},\vector{l},\nhat) & = & \measure
    \frac{F_{1}(x,\per{k};z,\per{q})F_{1}(1-y,-\per{l};1-z,-\per{q})}
         {F_{0}(x,\per{k};z,\per{q})} \ , \\
M_{3}^{(1)}(\vector{k},\vector{l},\nhat) & = & \measure
    \frac{F_{1}(x,\per{k};z,\per{q}) F_{1}(1-y,-\per{l};1-z,-\per{q})}
         {F_{2}(x,\per{k};z,\per{q}::1-y,-\per{l};1-z,-\per{q})} \ , \\
M_{4}^{(1)}(\vector{k},\vector{l},\nhat) & = & \measure
    \frac{F_{1}(1-x,-\per{k};1-z,-\per{q})F_{1}(1-y,-\per{l};1-z,-\per{q})}
         {F_{0}(x,\per{k};z,\per{q})} \ , \\
M_{5}^{(1)}(\vector{k},\vector{l},\nhat) & = & \measure
    \frac{F_{1}(1-x,-\per{k};1-z,-\per{q})F_{1}(y,\per{l};z,\per{q})}
         {F_{0}(x,\per{k};z,\per{q})} \ , \\
M_{6}^{(1)}(\vector{k},\vector{l},\nhat) & = & \measure
    \frac{F_{1}(1-x,-\per{k};1-z,-\per{q})F_{1}(y,\per{l};z,\per{q})}
         {F_{2}(1-x,-\per{k};1-z,-\per{q}::y,\per{l};z,\per{q})} \ , \label{eq:M_6}
\end{eqnarray}
where $F_{1}$ is defined in \eqn{eq:F1} and
\begin{eqnarray}
F_{0}(x,\per{k};z,\per{q}) & = &
    \frac{\per{k}^{2} + m^{2}}{x(1-x)} -
    \frac{\per{q}^{2} + m^{2}}{z(1-z)} \ , \label{eq:F0} \\
F_{2}(x,\per{k};z,\per{q}::y,\per{l};z',\per{q}') & = &
    \frac{\per{k}^{2} + m^{2}}{x} - \frac{\per{l}^{2} + m^{2}}{1-y} -
    \frac{(\per{k} - \per{q})^{2} + \mu^{2}}{x-z} -
    \frac{(\per{l} - \per{q}')^{2} + \mu^{2}}{y-z'} \ . \label{eq:F2}
\end{eqnarray}
Here,
\begin{equation}
\measure \equiv \int_{0}^{1} \frac{dz}{2z(1-z)} \int d^{2}\per{q} \ ,
   \label{eq:measure}
\end{equation}
and
\begin{eqnarray}
z & \equiv & \frac{q^{+}}{P^{+}}
    = \frac{1}{2} \left[ 1 + \frac{\ndot{q}}{\energy{q}} \right] \ , \\
\per{q} & = & \vector{q} - (\ndot{q})\nhat \ .
\end{eqnarray}
In the Appendix, we show explicitly the equivalence between the CVPT
and the sum of
$\tau$-ordered OFPT diagrams in \fig{fig:1,LT_OFPT}.
Since $(z;\per{q}) \longrightarrow (1-z;\per{q})$
as $\nhat \longrightarrow -\nhat$,
we have
\begin{eqnarray}
M_{4}^{(1)}(\vector{k},\vector{l},-\nhat) & = &
    M_{1}^{(1)}(\vector{k},\vector{l},\nhat) \ , \\
M_{5}^{(1)}(\vector{k},\vector{l},-\nhat) & = &
    M_{2}^{(1)}(\vector{k},\vector{l},\nhat) \ , \\
M_{6}^{(1)}(\vector{k},\vector{l},-\nhat) & = &
    M_{3}^{(1)}(\vector{k},\vector{l},\nhat) \ ,
\end{eqnarray}
and thus
\begin{eqnarray}
\tilde{M}_{1}^{(1)}(\vector{k},\vector{l}) & = &
    \tilde{M}_{4}^{(1)}(\vector{k},\vector{l}) \ , \label{eq:T_M_1} \\
\tilde{M}_{2}^{(1)}(\vector{k},\vector{l}) & = &
    \tilde{M}_{5}^{(1)}(\vector{k},\vector{l}) \ , \\
\tilde{M}_{3}^{(1)}(\vector{k},\vector{l}) & = &
    \tilde{M}_{6}^{(1)}(\vector{k},\vector{l}) \ , \label{eq:T_M_3}
\end{eqnarray}
where
\begin{equation}
\tilde{M}_{i}^{(1)}(\vector{k},\vector{l}) =
\frac{1}{4\pi} \int d\nhat M_{i}^{(1)}(\vector{k},\vector{l},\nhat) \ .
\end{equation}

Again, we may summarize our results for the next-to-leading order ladder diagrams
as follows:
\begin{equation}
M_{\rm SUM}^{(1)}(\vector{k},\vector{l})
 = \sum_{i=1}^{6} M_{i}^{(1)}(\vector{k},\vector{l},\nhat)
 = \sum_{i=1}^{6} \tilde{M}_{i}^{(1)}(\vector{k},\vector{l})
 = 2 \sum_{i=1}^{3} \tilde{M}_{i}^{(1)}(\vector{k},\vector{l}) \ .
\end{equation}
Thus, we need to calculate only the three (not six) diagrams to obtain
$M_{\rm SUM}^{(1)}(\vector{k},\vector{l})$. In the next section,
we calculate numerically $\tilde{M}_{i}^{(1)}, (i=1,2,\cdots,6)$ and
verify Eqs. (\ref{eq:T_M_1})-(\ref{eq:T_M_3}).
Our numerical results also show
how small the higher Fock-state contribution
$\tilde{M}_{3}^{(1)}(\vector{k},\vector{l})$ $\left(\tilde{M}_{6}^{(1)}(\vector{k},\vector{l})\right)$ is.

\section{Numerical Results}\label{sec:SEC_num...}

As shown explicitly by $\nhat$-dependence, each amplitude in the
equal-$\tau$ OFPT does not have rotational symmetry. Nevertheless, all
the $\nhat$-dependence from each amplitude must cancel each other
if we sum them up. The rotational symmetry must be recovered in the
Feynman amplitude level. We first confirm this numerically using
$M_{i}^{(1)}(\vector{k},\vector{l},\nhat)$ given
by Eqs. (\ref{eq:M_1})-(\ref{eq:M_6}).
For the numerical calculation, we first observe that the amplitudes
$M_{i}^{(1)}$ are complex in general. We thus separate the real and
imaginary parts of $M_{i}^{(1)}$ using the usual relation
\begin{equation}
\lim_{\epsilon \longrightarrow 0} \frac{1}{x + i\epsilon}
 = {\rm PV}\left(\frac{1}{x}\right) - i\pi\delta(x) \ ,
\end{equation}
where ${\rm PV}\left(\frac{1}{x}\right)$ is the principle value of $\frac{1}{x}$.
Hence the real part of $M_{i}^{(1)}(\vector{k},\vector{l},\nhat), \ i=1,2,4,5$
are given by Cauchy principle values. However, the higher Fock-state contributions
$M_{3}^{(1)}$ and $M_{6}^{(1)}$ turn out to be real because the intermediate
state of higher Fock-states cannot go to the on-energy-shell.
In this numerical work, we will focus only on the real part of each amplitude.
For the Cauchy principle value calculation, we change the integration
variables, ($z,\per{q}$), into ($\vector{q}$) with the fixed $\nhat$ and do the integration
over a spherical coordinate of $\vector{q}$.
Since
\begin{equation}
z = \frac{1}{2} \left[ 1 + \frac{\ndot{q}}{\energy{q}} \right]
\end{equation}
and $\vector{q}^{2} = (\ndot{q})^{2} + \per{q}^{2}$,
one can obtain
\begin{equation}
\frac{dz}{2z(1-z)} = \frac{d(\ndot{q})}{\energy{q}} \ ,
\end{equation}
and thus the integration measure defined in \eqn{eq:measure} can be rewritten as
\begin{equation}
\measure = \int \frac{d^{3}\vector{q}}{\energy{q}} \ ,
\end{equation}
where $d^{3}\vector{q} = \vector{q}^{2}d|\vector{q}|d\Omega(\vector{q})$.
Using the relations between the variable sets
$(x,\per{k};y,\per{l};z,\per{q})$ and $(\vector{k},\vector{l},\vector{q})$
with the fixed $\nhat$,
one can change the functions $F_{0}, F_{1}$ and $F_{2}$ given by Eqs.
(\ref{eq:F0}), (\ref{eq:F1}) and (\ref{eq:F2}), respectively,
as follows:
\begin{eqnarray}
F_{0}(x,\per{k};z,\per{q})
 & = & 4(\vector{k}^{2} - \vector{q}^{2}) \ , \\
F_{1}(x,\per{k};z,\per{q}) & = &
F_{1}(\vector{k},\nhat;|\vector{q}|,\Omega(\vector{q})) \ , \\
F_{2}(x,\per{k};z,\per{q}::z',\per{q}';y,\per{l}) & = &
F_{2}(\vector{k},\vector{l},\nhat;|\vector{q}|,\Omega(\vector{q});|\vector{q}'|,\Omega(\vector{q}')) \ .
\end{eqnarray}
Also, for the numerical calculation of a Cauchy principle value(PV),
we note that for $x_{0} > 0$
\begin{equation}
{\rm PV}\int_{0}^{\infty} \frac{f(x)}{x^{2} - x_{0}^{2}} =
    \int_{0}^{\infty}\frac{f(x) - f(x_{0})}{x^{2} - x_{0}^{2}} \ .
\end{equation}
Thus the real part of $M_{1}^{(1)}(\vector{k},\vector{l},\nhat)$ is given by
\begin{eqnarray}
{\rm Re}\left\{ M_{1}^{(1)}(\vector{k},\vector{l},\nhat) \right\}
    & = & {\rm PV} \measure
    \frac{F_{1}(x,\per{k};z,\per{q})F_{1}(y,\per{l};z,\per{q})}
         {F_{0}(x,\per{k};z,\per{q})} \\
    & = & \int d\Omega(\vector{q}) \left[
    \int_{0}^{\infty} d|\vector{q}|
    \frac{F(\vector{k},\vector{l},\nhat;|\vector{q}|,\Omega(\vector{q}))
        - F(\vector{k},\vector{l},\nhat;|\vector{k}|,\Omega(\vector{q}))}
         {\vector{k}^{2} - \vector{q}^{2}} \right] \ ,
\end{eqnarray}
where
\begin{equation}
F(\vector{k},\vector{l},\nhat;|\vector{q}|,\Omega(\vector{q}))
= \frac{\vector{q}^{2}}{4\energy{q}}
  F_{1}(\vector{k},\nhat;|\vector{q}|,\Omega(\vector{q}))
  F_{1}(\vector{l},\nhat;|\vector{q}|,\Omega(\vector{q})) \ .
\end{equation}
The real parts of all other amplitudes
can be written similarly.

\section{Numerical Results}\label{sec:SEC_con...}

For the explicit example of numerical results,
we choose the following kinematics without any loss of generality;
\begin{eqnarray}
\vector{k} & = & |\vector{k}| (0, 0, 1) \ , \\
\vector{l} & = & |\vector{k}| (0, \sin\Theta, \cos\Theta) \ , \\
\nhat & = & (\sin\theta_{n}\cos\phi_{n}, \sin\theta_{n}\sin\phi_{n}, \cos\theta_{n}) \ ,
\end{eqnarray}
where $\Theta$ is an angle between $\vector{k}$ and $\vector{l}$, and
$\theta_{n}(\phi_{n})$ is a polar(azimuthal) angle of $\nhat$.

Because we are interested in the dependence of the scattering amplitude
on the direction $\nhat$, we fix the scattering plane as the plane made by
$\hat{\vector{y}}$ and $\hat{\vector{z}}$ and the direction of initial
momentum $\vector{k}$ as $\hat{\vector{z}}$ and then vary the
direction $\nhat$. The effect of rotating the direction $\nhat$
in a given scattering plane defined by its perpendicular direction
$\vector{k}\times\vector{l}$ is equivalent to the effect
of rotating $\vector{k}\times\vector{l}$ in a given direction of the
lightcone time evolution, {\it e.g.}, $\tau = t + z$. In any case,
the point is the dynamics dependent on the relative angle between
$\nhat$ and $\vector{k}\times\vector{l}$\cite{ref:JKM}.

In Figs.~\ref{fig:EACH__0_1}--\ref{fig:EACH__1_1}, the scattering amplitudes of each diagram are plotted for
$|\vector{k}| = 1.0$ and $\mu = 1.0$ in units of a mass of scattering particle, $m$,
with given scattering angle, $\Theta = 0, \pi/6, \pi/4, \pi/3, \pi/2, 2\pi/3, 3\pi/4, 5\pi/6$ and $\pi$.
From these figures, we can easily see that each amplitude has the dependence
on the angles of $\nhat$, $\theta_{n}$ and $\phi_{n}$, but the sum of all
amplitudes $M_{\rm SUM}^{(1)}$ is independent from $\theta_{n}$ and $\phi_{n}$
within the numerical error. This shows the recovery of rotational symmetry
in the Feynman amplitude level\cite{ref:Bakker_I}.
It is also very interesting to note that
the higher Fock-state contributions, $M_{3}^{(1)}$ and $M_{6}^{(1)}$,
are quite suppressed\cite{ref:Bakker_II}.
The similar behavior has been observed for various
scattering angle $\Theta$.
The real part numerical values of $\tilde{M}_{i}^{(1)}$ are
listed in \tab{tab:n_average} for various $\Theta$
with given $|\vector{k}|/m=1.0,\mu/m=1.0$.
The \tab{tab:n_average} also
verifies $\tilde{M}_{1}^{(1)}=\tilde{M}_{4}^{(1)},
\tilde{M}_{2}^{(1)}=\tilde{M}_{5}^{(1)}$ and
$\tilde{M}_{3}^{(1)}=\tilde{M}_{6}^{(1)}$
for various $\Theta$ within the numerical errors.
Finally in \tab{tab:s_wave}, the $S$-wave scattering amplitude
given by
\begin{equation}
\swave{i} \equiv \frac{1}{4\pi} \int_{0}^{\pi} \sin\Theta d\Theta \average{i} 
\end{equation}
is listed for various $|\vector{k}|/m=1.0,0.1,0.01$ with given $\mu/m=1.0$.
This also numerically verifies the smallness of the higher Fock-state contributions.

\section{Conclusions and Discussions}
\begin{figure}[ht]
	\centerline{\epsffile{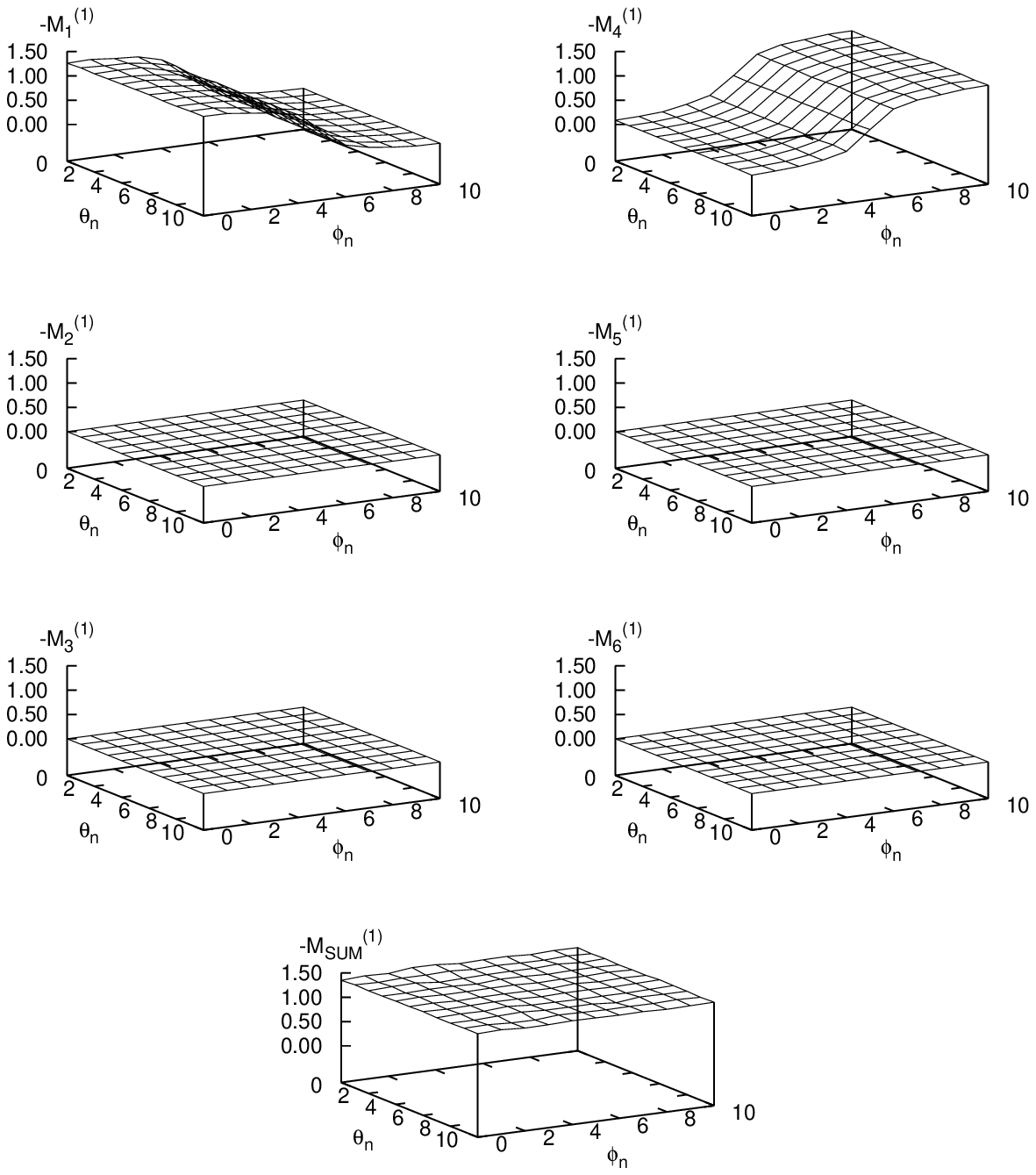}}
        \caption[]{
            The scattering amplitude of each diagram for $\Theta=0$.
            $\theta_{n}$-axis and $\phi_{n}$-axis are scaled
            in units of $\pi/10$ and in units of $\pi/5$, respectively.
        }
        \label{fig:EACH__0_1}
    \end{figure}

\begin{figure}[ht]
	\centerline{\epsffile{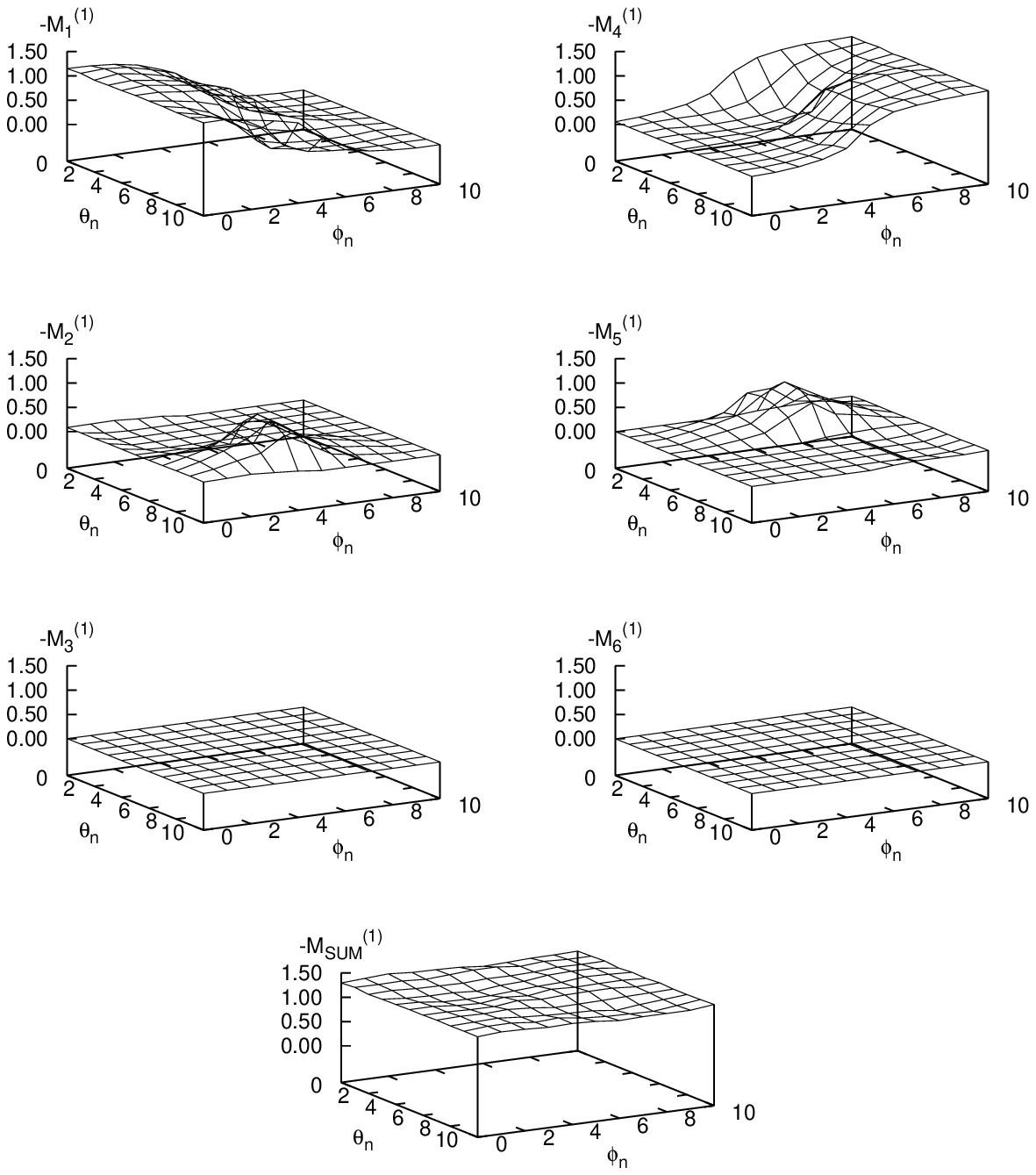}}
        \caption[]{
            The scattering amplitude of each diagram for $\Theta=\pi/6$.
            $\theta_{n}$-axis and $\phi_{n}$-axis are scaled
            in units of $\pi/10$ and in units of $\pi/5$, respectively.
        }
        \label{fig:EACH__1_6}
    \end{figure}

\begin{figure}[ht]
	\centerline{\epsffile{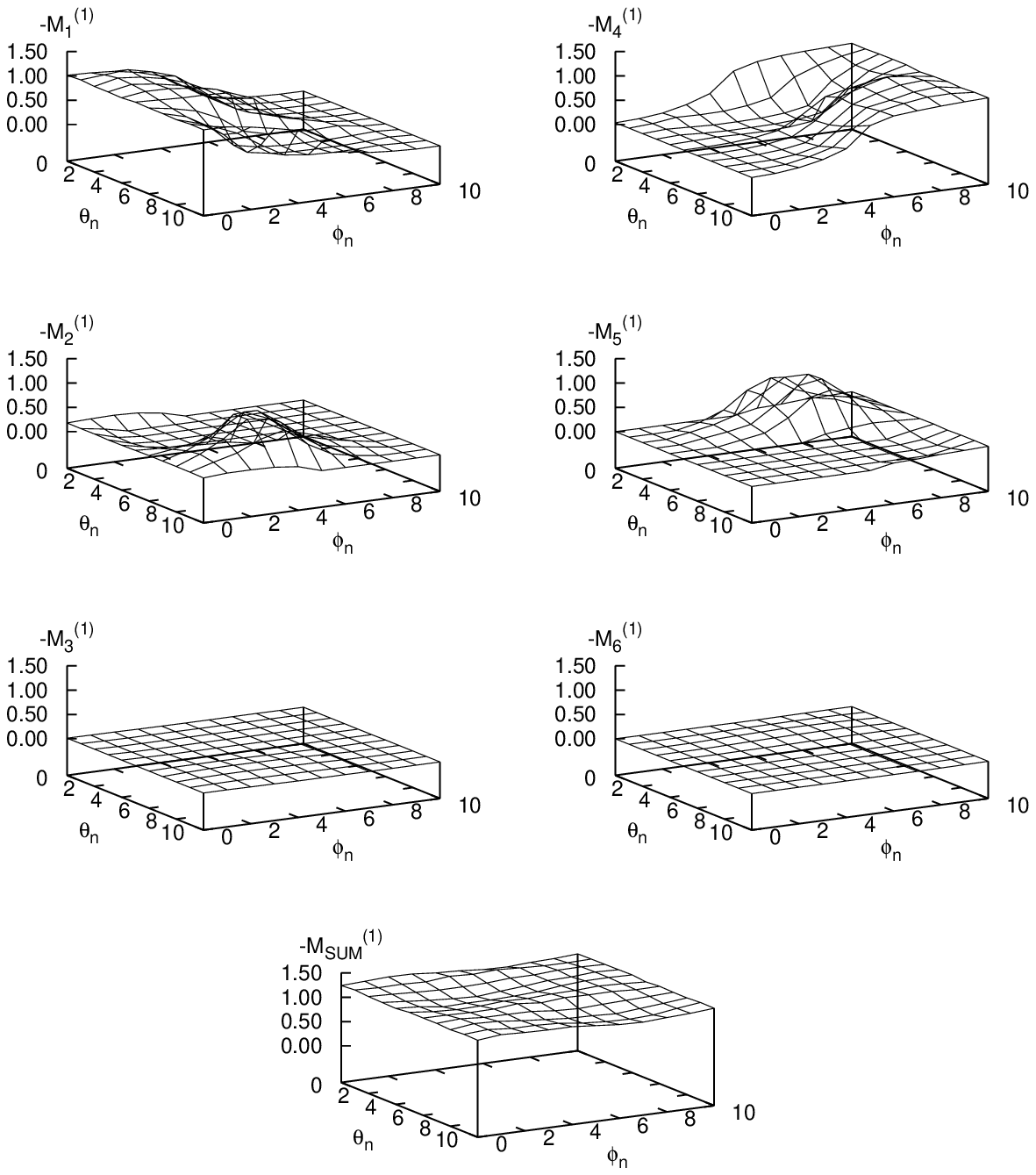}}
        \caption[]{
            The scattering amplitude of each diagram for $\Theta=\pi/4$.
            $\theta_{n}$-axis and $\phi_{n}$-axis are scaled
            in units of $\pi/10$ and in units of $\pi/5$, respectively.
        }
        \label{fig:EACH__1_4}
    \end{figure}

\begin{figure}[ht]
	\centerline{\epsffile{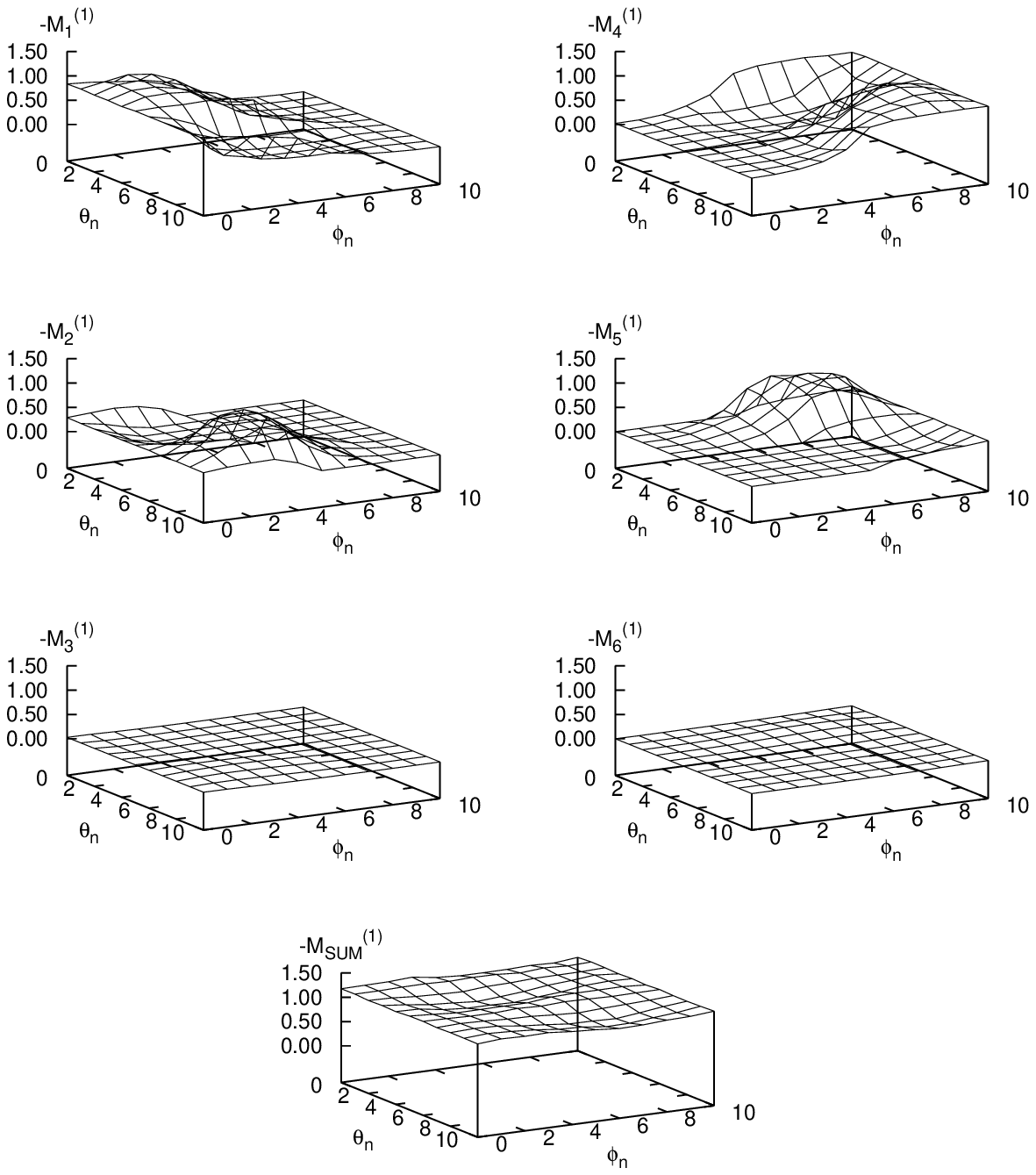}}
        \caption[]{
            The scattering amplitude of each diagram for $\Theta=\pi/3$.
            $\theta_{n}$-axis and $\phi_{n}$-axis are scaled
            in units of $\pi/10$ and in units of $\pi/5$, respectively.
        }
        \label{fig:EACH__1_3}
    \end{figure}

\begin{figure}[ht]
	\centerline{\epsffile{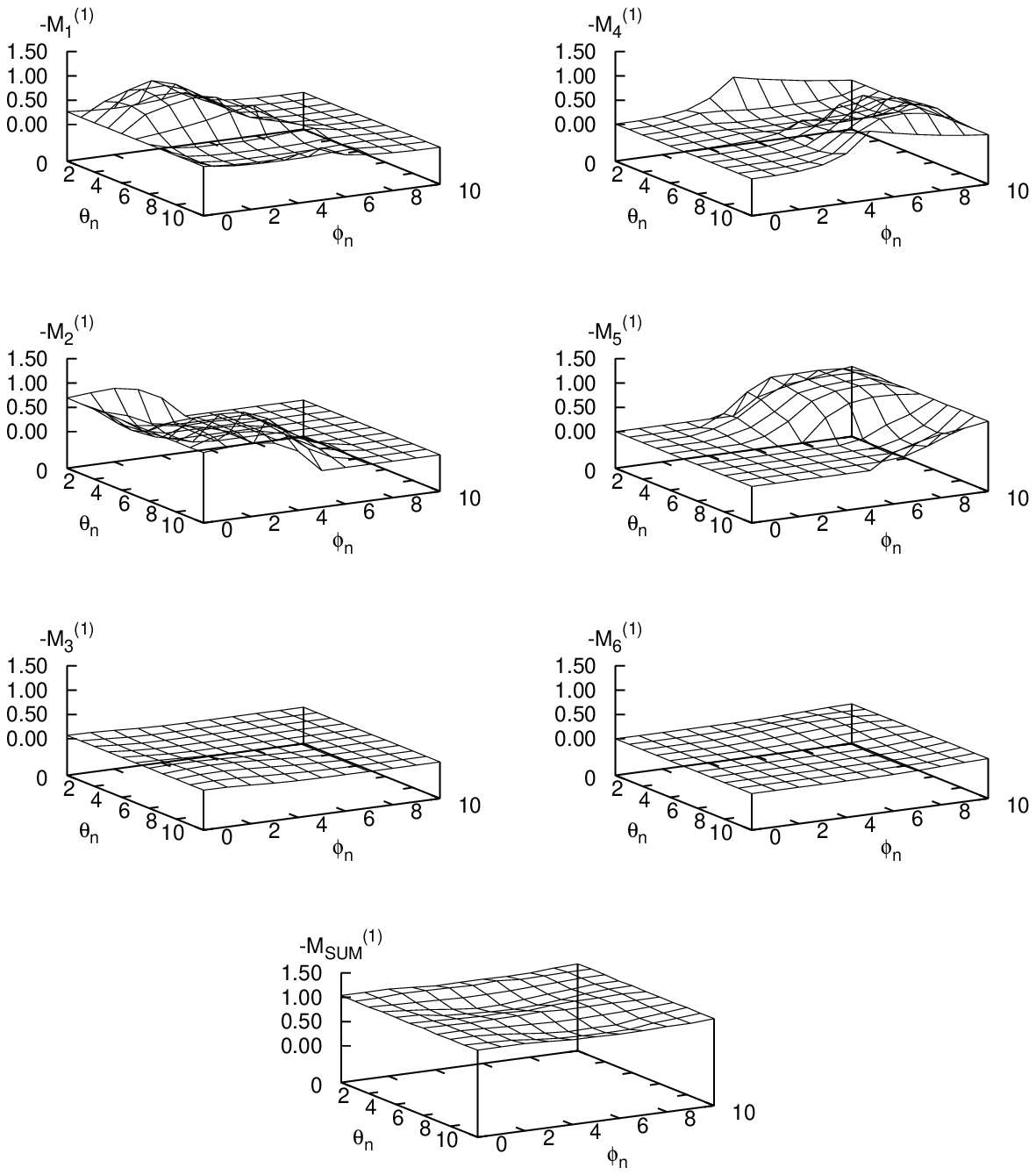}}
        \caption[]{
            The scattering amplitude of each diagram for $\Theta=\pi/2$.
            $\theta_{n}$-axis and $\phi_{n}$-axis are scaled
            in units of $\pi/10$ and in units of $\pi/5$, respectively.
        }
        \label{fig:EACH__1_2}
    \end{figure}

\begin{figure}[ht]
	\centerline{\epsffile{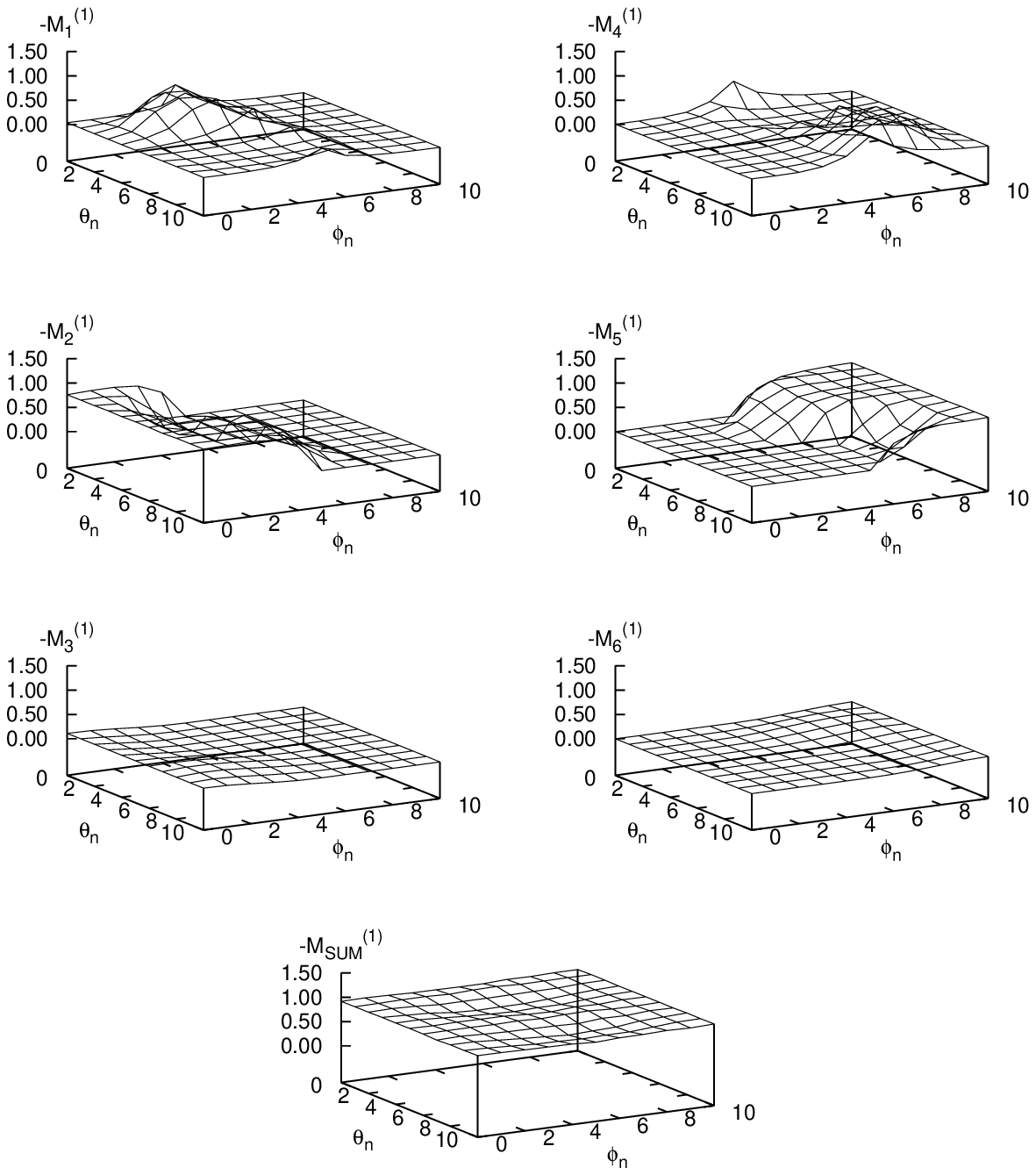}}
        \caption[]{
            The scattering amplitude of each diagram for $\Theta=2\pi/3$.
            $\theta_{n}$-axis and $\phi_{n}$-axis are scaled
            in units of $\pi/10$ and in units of $\pi/5$, respectively.
        }
        \label{fig:EACH__2_3}
    \end{figure}

\begin{figure}[ht]
	\centerline{\epsffile{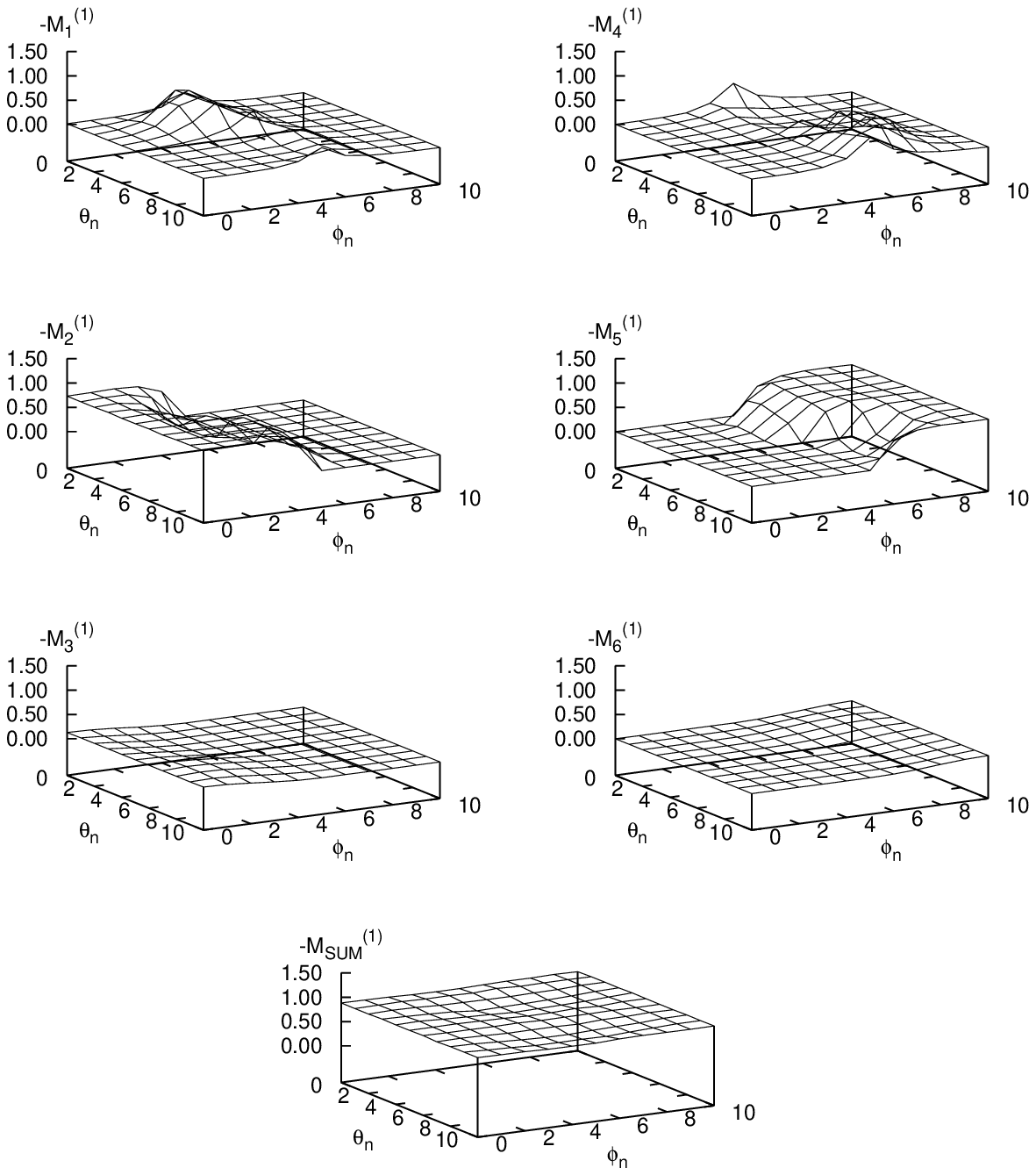}}
        \caption[]{
            The scattering amplitude of each diagram for $\Theta=3\pi/4$.
            $\theta_{n}$-axis and $\phi_{n}$-axis are scaled
            in units of $\pi/10$ and in units of $\pi/5$, respectively.
        }
        \label{fig:EACH__3_4}
    \end{figure}

\begin{figure}[ht]
	\centerline{\epsffile{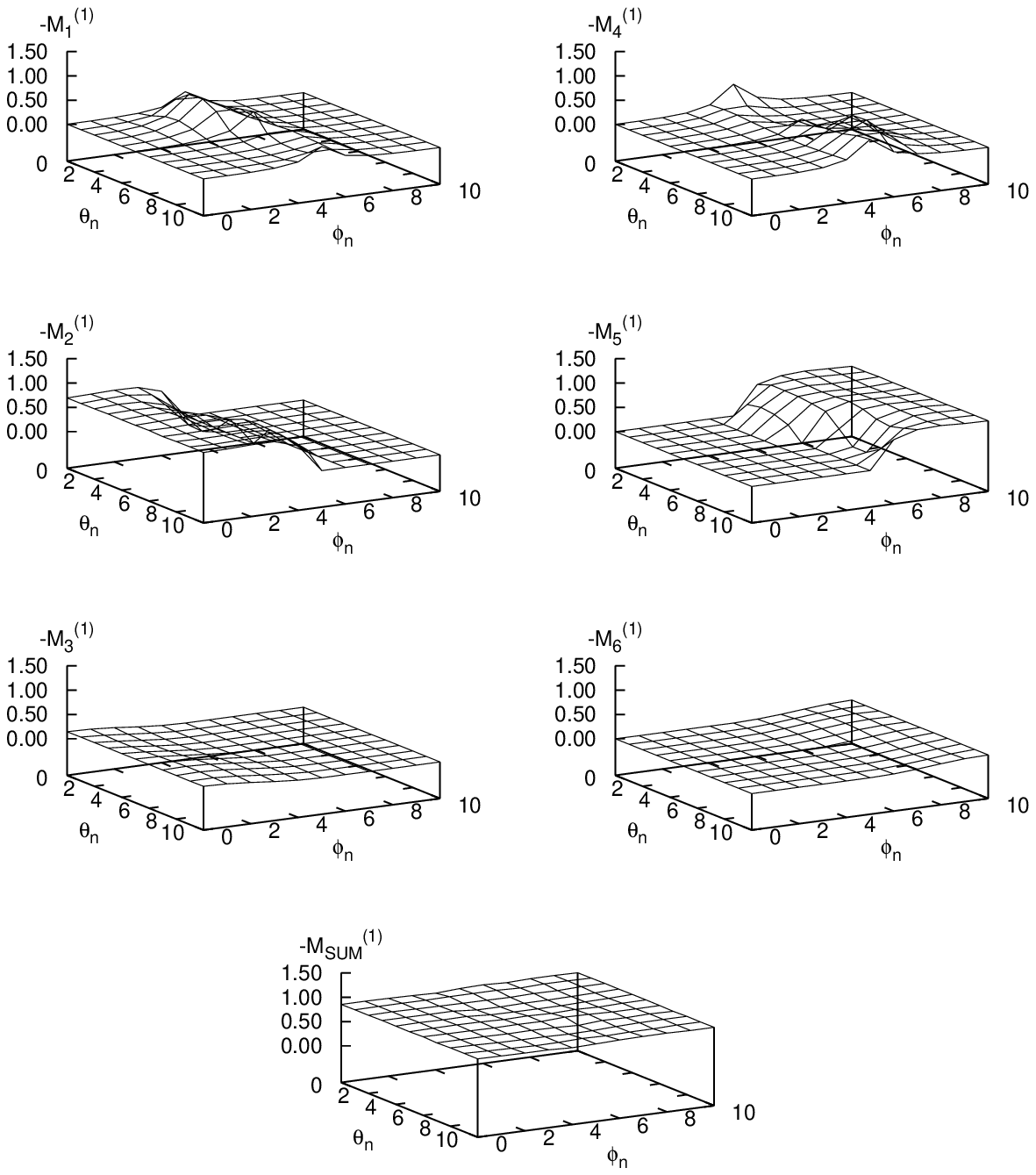}}
        \caption[]{
            The scattering amplitude of each diagram for $\Theta=5\pi/6$.
            $\theta_{n}$-axis and $\phi_{n}$-axis are scaled
            in units of $\pi/10$ and in units of $\pi/5$, respectively.
        }
        \label{fig:EACH__5_6}
    \end{figure}

\begin{figure}[ht]
	\centerline{\epsffile{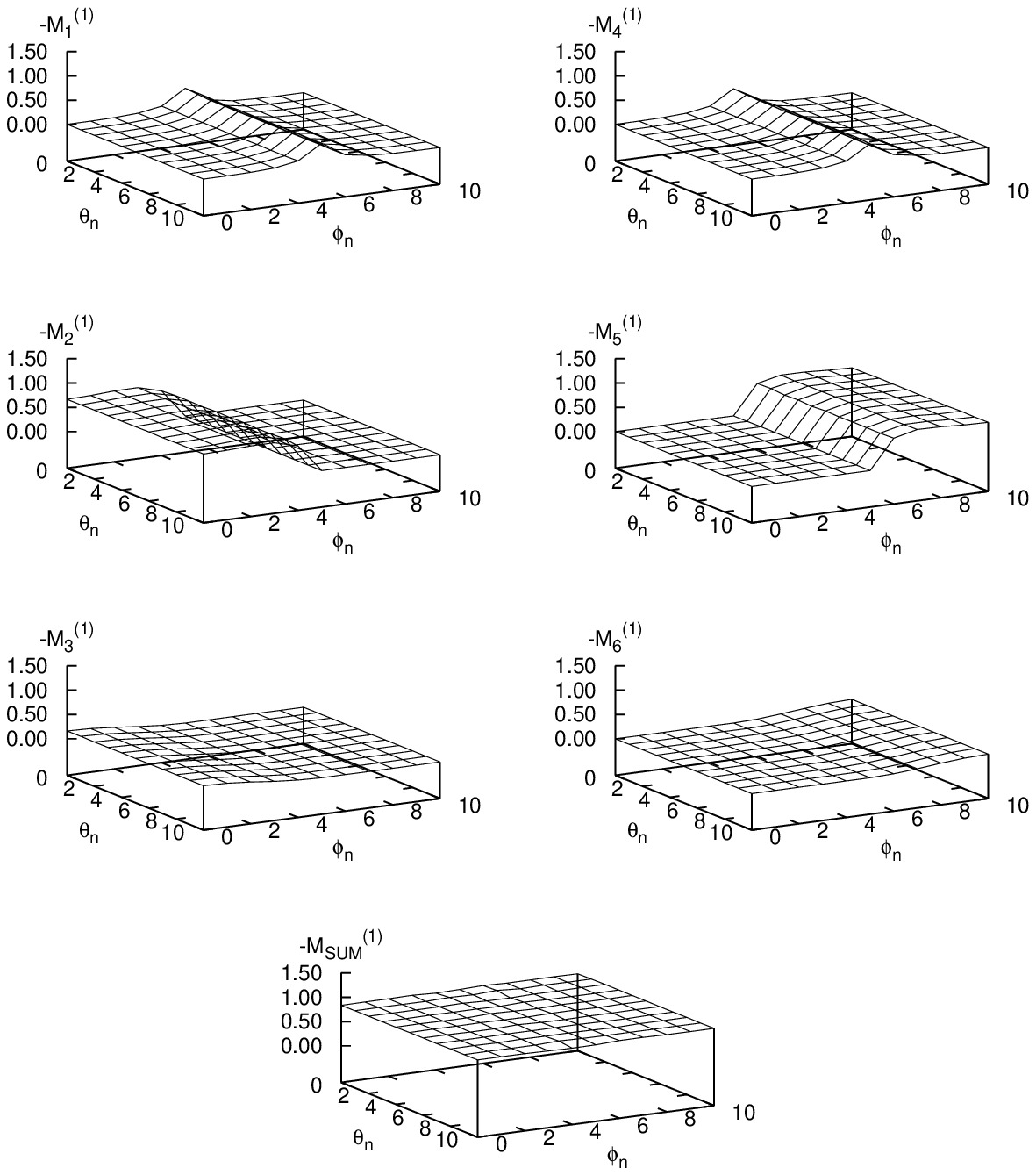}}
        \caption[]{
            The scattering amplitude of each diagram for $\Theta=\pi$.
            $\theta_{n}$-axis and $\phi_{n}$-axis are scaled
            in units of $\pi/10$ and in units of $\pi/5$, respectively.
        }
        \label{fig:EACH__1_1}
    \end{figure}

In this work, we have shown that each $\tau$-ordered amplitude can be made as
the Lorentz invariant amplitude by taking advantage of a distinguished feature
in the lightcone quantization and making an average over the lightcone surface
defined by $\nhat$. Such process of averaging was possible in the lightcone
quantization method because the rotation which is the dynamical
part of this quantization method is actually compact. This feature
is drastically different from the ordinary equal-$t$ quantization, where
the dynamical part occurs in the boost operation but the parameter
space of this operation is not closed.
We regard this as an explicit example of advantage in the equal-$\tau$
quantization over the equal-$t$ quantization.
The rotation average of each $\tau$-ordered scattering amplitude
not only provided the Lorentz-invariant assessment of each amplitude but also
reduced the number of diagrams to be calculated
if the masses of two bodies are same.
For the explicit numerical examples, we have calculated the real part of
next-to-leading order ladder diagrams in $\phi^{3}$ theory. As shown in
Figs.~\ref{fig:EACH__0_1}--\ref{fig:EACH__1_1}, the sum of all diagrams
is always independent of the $\nhat$ choice, {\it i.e.}, reference-frame
independent or Lorentz invariant, even though each $\tau$-ordered diagram
($M_{i}^{(1)}, i=1,2,\cdots,6$) is not Lorentz invariant. Also, the
numerical values of $\nhat$-averaged amplitudes
(${\rm Re}\left\{\tilde{M}_{i}^{(1)}\right\}, i=1,2,\cdots,6$) presented in \tab{tab:n_average} not
only verify the equivalence, $\tilde{M}_{1}^{(1)} = \tilde{M}_{4}^{(1)}$, etc.,
but also show the significant suppression of
$\tilde{M}_{3}^{(1)}$ ($\tilde{M}_{6}^{(1)}$)
compare to 
$\tilde{M}_{1}^{(1)}$ ($\tilde{M}_{4}^{(1)}$) or
$\tilde{M}_{2}^{(1)}$ ($\tilde{M}_{5}^{(1)}$)
for whole range of scattering angle.
The $S$-wave scattering amplitudes $\tilde{M}_{i,S}^{(1)}$ presented in
\tab{tab:s_wave} also verify the negligible contribution from the
higher Fock-state intermediate states. Thus $\nhat$-averaging process
exhibits a unique advantage of assessing the contribution from each
intermediate Fock-states.
This brings up further interesting application to the gauge theory such as
QED and QCD as well as to the multi-body scattering amplitudes as future works.

\centerline{\bf ACKNOWLEDGEMENT}

CRJ is very grateful to the CTP of SNU and the Asia Pacific Center for
Theoretical Physics for their warm hospitality during his stay while this
work was completed. The work of GHK and DPM was supported in part by
KOSEF through CTP, SNU and in part by the Korea MOE(BRSI-97-2418).
CRJ was supported in part by the U.S. DOE under contracts DE-FG02-96ER40947.

\renewcommand{\theequation}{A-\arabic{equation}}
\setcounter{equation}{0}
\centerline{\bf APPENDIX}

In this appendix, we show the equivalence between the CVPT diagram in
\fig{fig:1,CVPT} and the sum of $\tau$-ordered OFPT diagrams in \fig{fig:1,LT_OFPT}.
The method used in this Appendix was presented in Ref.\cite{ref:BJS} for
the Bethe-Salpeter approach. Recently, Ligterink and Bakker\cite{ref:LB}
also proposed a general algorithm that produces the $\tau$-ordered diagrams
from any Feynman diagram.
The scattering amplitude from the diagram in \fig{fig:1,CVPT}
is given by
\begin{eqnarray*}
M^{1}(\vector{k},\vector{l}) = \int \frac{d^{4}q}{(2\pi)^{4}}
 & & \frac{1}{q^{2} - m^{2} + i\epsilon}
     \frac{1}{(k-q)^{2} - \lambda^{2} + i\epsilon} \\
 & & \frac{1}{(P-q)^{2} - m^{2} + i\epsilon}
     \frac{1}{(q-l)^{2} - \lambda^{2} + i\epsilon} \ .
\end{eqnarray*}

In terms of lightcone variables,
$M^{(1)}(\vector{k},\vector{l})$ can be rewritten as
\begin{eqnarray}
M^{(1)}(\vector{k},\vector{l},\nhat) = \int_{0}^{P^{+}}\frac{dq^{+}}{2(2\pi)}
\int\frac{d^{2}\per{q}}{(2\pi)^{2}} \int_{-\infty}^{\infty}\frac{dq^{-}}{2\pi}
 & & \frac{1}{q^{+}q^{-} - \per{q}^{2} - m^{2} + i\epsilon} \nonumber \\
 & & \frac{1}{(k^{+} - q^{+})(k^{-} - q^{-}) - (\per{k} - \per{q})^{2}
	- \lambda^{2} + i\epsilon} \nonumber \\
 & & \frac{1}{(P^{+} - q^{+})(P^{-} - q^{-}) - \per{q}^{2} - m^{2} + i\epsilon} \nonumber \\
 & & \frac{1}{(q^{+} - l^{+})(q^{-} - l^{-}) - (\per{q} - \per{l})^{2}
	- \lambda^{2} + i\epsilon} \ .
\end{eqnarray}

If we define momentum fractions, $x,y,z$ as
\begin{eqnarray*}
k^{+} &\equiv& xP^{+} \ , \\
l^{+} &\equiv& yP^{+} \ , \\
q^{+} &\equiv& zP^{+} \ ,
\end{eqnarray*}
and make a change of variable such as $P^{+}q^{-} \longrightarrow q^{-}$,
we obtain
\begin{eqnarray}
M^{(1)}(\vector{k},\vector{l},\nhat)
 & = & \int_{0}^{1}\frac{dz}{2(2\pi)z(1-z)}
\frac{1}{x-z} \frac{1}{z-y}
\int\frac{d^{2}\per{q}}{(2\pi)^{2}}
\nonumber \\
 &   & \int_{-\infty}^{\infty}\frac{dq^{-}}{2\pi}
       \frac{1}{q^{-} - q_{1}^{-}} \frac{1}{q^{-} - q_{2}^{-}}
       \frac{1}{q^{-} - q_{3}^{-}} \frac{1}{q^{-} - q_{4}^{-}} \ ,
\end{eqnarray}
where
\begin{eqnarray*}
q_{1}^{-} & = & \frac{\per{q}^{2} + m^{2}}{z} - i\frac{\epsilon}{z} \ , \\
q_{2}^{-} & = & P^{+}k^{-} - \frac{(\per{k} - \per{q})^{2} + \lambda^{2}}{x - z}
		+ i\frac{\epsilon}{x - z} \ , \\
q_{3}^{-} & = & P^{+}P^{-} - \frac{\per{q}^{2} + m^{2}}{1 - z}
		+ i\frac{\epsilon}{1 - z} \ , \\
q_{4}^{-} & = & P^{+}l^{-} + \frac{(\per{q} - \per{l})^{2} + \lambda^{2}}{z - y}
		- i\frac{\epsilon}{z - y} \ .
\end{eqnarray*}

Here, the on-shell condition, $k^{2}=m^{2}$ gives
\begin{equation}
P^{+}k^{-} = \frac{\per{k}^{2} + m^{2}}{x} \ ,
\end{equation}
and similarly $l^{2}=m^{2}$ gives
\begin{equation}
P^{+}l^{-} = \frac{\per{l}^{2} + m^{2}}{y} \ .
\end{equation}
Also, from the zero binding energy of the initial and final scattering
particles, we can get
\begin{equation}
P^{+}P^{-} = \frac{\per{k}^{2} + m^{2}}{x(1-x)}
           = \frac{\per{l}^{2} + m^{2}}{y(1-y)} \ .
\end{equation}
Now, if we introduce the notation $(i,j)$ for $i \neq j$ ($i,j=1,2,3,4$) which
is defined as
\begin{equation}
(i,j) \equiv \frac{1}{q_{i}^{-} - q_{j}^{-}} \ ,
\end{equation}
then the following properties are obtained:
\begin{eqnarray}
(i,j)      & = & -(j,i) \ , \\
(i,j)(i,k) & = & (i,j)(j,k) - (i,k)(j,k) \ .
\end{eqnarray}
With this notation, the functions, $F_{i}$
needed for the $\tau$-ordered amplitudes $M_{i}^{(1)}$ in
Eqs.~(\ref{eq:M_1})-(\ref{eq:M_6}) are given by
\begin{eqnarray}
F_{0}(x,\per{k};z,\per{q})       & = & \frac{1}{(3,1)} \ , \\
F_{1}(x,\per{k};z,\per{q})       & = & \frac{\theta(x-z)}{x-z} \ (2,1) \ , \\
F_{1}(1-x,-\per{k};1-z,-\per{q}) & = & \frac{\theta(z-x)}{z-x} \ (3,2) \ , \\
F_{1}(y,\per{l};z,\per{q})       & = & \frac{\theta(y-z)}{y-z} \ (4,1) \ , \\
F_{1}(1-y,-\per{l};1-z,-\per{q}) & = & \frac{\theta(z-y)}{z-y} \ (3,4) \ , \\
F_{2}(x,\per{k};z,\per{q}::1-y,-\per{l};1-z,-\per{q}) & = & \frac{1}{(2,4)} \ , \\
F_{2}(1-x,-\per{k};1-z,-\per{q}::y,\per{l};z,\per{q}) & = & \frac{1}{(4,2)} \ .
\end{eqnarray}

In case of $x>z,y>z$, we have one pole($q_{1}^{-}$) in the lower half plane and
three poles($q_{2}^{-},q_{3}^{-},q_{4}^{-}$) in the upper plane. Hence if we do
a contour integration by enclosing a lower half plane,
we obtain the contribution
of integral as
\begin{equation}
i \int_{0}^{1}\frac{dz}{2(2\pi)z(1-z)} \frac{\theta(x-z)}{x-z}
\frac{\theta(y-z)}{y-z} \int\frac{d\per{q}}{(2\pi)^{2}}
\ (1,2)(1,3)(1,4) \ ,
\end{equation}
which is equal to $-i/(2\pi)^{3} M_{1}^{(1)}(\vector{k},\vector{l},\nhat)$.

In case of $x>z,y<z$, we have two poles($q_{2}^{-},q_{3}^{-}$) in the upper half
plane and two poles($q_{1}^{-},q_{4}^{-}$) in the lower half plane.
Doing a contour integration by enclosing a lower half plane, the contribution of
integral becomes
\begin{equation}
-i \int_{0}^{1}\frac{dz}{2(2\pi)z(1-z)} \frac{\theta(x-z)}{x-z}
\frac{\theta(z-y)}{z-y} \int\frac{d\per{q}}{(2\pi)^{2}}
\left[ (1,2)(1,3)(1,4) + (4,1)(4,2)(4,3) \right] \ . \label{eq:A}
\end{equation}
Since
\begin{eqnarray*}
(1,2)(1,3)(1,4) + (4,1)(4,2)(4,3)
 & = & (1,2) [ (1,3)(3,4) - (1,4)(3,4) ] + (4,1)(4,2)(4,3) \\
 & = & (2,1)(3,1)(3,4) + [ (2,1)(1,4) - (2,4)(1,4) ] (3,4) \\
 & = & (2,1)(3,1)(3,4) + (2,1)(2,4)(3,4) \ ,
\end{eqnarray*}
\eqn{eq:A} is equal to
$-i/(2\pi)^{3} ( M_{2}^{(1)}(\vector{k},\vector{l},\nhat)
               + M_{3}^{(1)}(\vector{k},\vector{l},\nhat) )$.
Similarly, the equivalence between the remaining cases of pole positions
and the rest of the $\tau$-ordered diagrams can be shown by similar steps
of $q^{-}$ contour integration. This shows that the sum of six $\tau$-ordered
diagrams in \fig{fig:1,LT_OFPT} is same as the single Feynman digram
in \fig{fig:1,CVPT}.

\newpage

\begin{table}
{\center
  \begin{tabular}{|r||r|r|r|r|r|r|r|}
    \hline
    $\Theta$   &$-\average{1}$&$-\average{2}$&$-\average{3}$&$-\average{4}$&$-\average{5}$&$-\average{6}$&$-{\rm Re}\left\{M_{\rm SUM}^{(1)}\right\}$ \\ \hline\hline
    $      0 $ & 0.6870 & 0.0000 & 0.0000 & 0.6858 & 0.0000 & 0.0000 & 1.3728 \\
    $  \pi/6 $ & 0.4906 & 0.1488 & 0.0050 & 0.4899 & 0.1492 & 0.0050 & 1.2884 \\
    $  \pi/4 $ & 0.4039 & 0.1907 & 0.0089 & 0.4036 & 0.1902 & 0.0089 & 1.2062 \\
    $  \pi/3 $ & 0.3315 & 0.2156 & 0.0125 & 0.3326 & 0.2159 & 0.0125 & 1.1207 \\
    $  \pi/2 $ & 0.2263 & 0.2417 & 0.0186 & 0.2258 & 0.2420 & 0.0186 & 0.9731 \\
    $ 2\pi/3 $ & 0.1599 & 0.2591 & 0.0231 & 0.1602 & 0.2588 & 0.0231 & 0.8842 \\
    $ 3\pi/4 $ & 0.1362 & 0.2690 & 0.0248 & 0.1366 & 0.2689 & 0.0247 & 0.8602 \\
    $ 5\pi/6 $ & 0.1175 & 0.2797 & 0.0260 & 0.1163 & 0.2789 & 0.0260 & 0.8444 \\
    $    \pi $ & 0.0956 & 0.2954 & 0.0271 & 0.0953 & 0.2959 & 0.0270 & 0.8362 \\
    \hline
  \end{tabular}
  \caption{
    The real part contributions of each $\hat{\vector{n}}$--averaged
    scattering amplitudes for various $\Theta$
    with fixed $\lambda=1.0m, |\vector{k}|=1.0m$.
  }
  \label{tab:n_average}
}
\end{table}

\begin{table}
\center{
  \begin{tabular}{|r||r|r|r|r|r|r|}
    \hline
    $|\vector{k}|/m$ &$-\swave{1}$&$-\swave{2}$&$-\swave{3}$&$-\swave{4}$&$-\swave{5}$&$-\swave{6}$ \\ \hline\hline
    $0.1$         & 1.5250 & 0.3171 & 0.0008 & 1.5247 & 0.3144 & 0.0008 \\
    $1.0$         & 0.2585 & 0.2307 & 0.0173 & 0.2592 & 0.2303 & 0.0173 \\
    $ 10$         & 0.0063 & 0.0036 & 0.0002 & 0.0053 & 0.0035 & 0.0002 \\
    \hline
  \end{tabular}
  \caption{
    The real part contributions of each diagram for both $\Theta$ and
    $\hat{\vector{n}}$--average of a scattering amplitude
    for various $|\vector{k}|$ with fixed $\lambda=1.0m$.
  }
  \label{tab:s_wave}
}
\end{table}

\newpage

\begin{figure}[ht]
\centerline{\epsffile{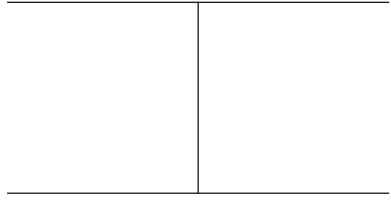}}
\caption[]{
The lowest diagram for a scattering amplitude in CVPT.
}
\label{fig:0,CVPT}
\end{figure}

\begin{figure}[ht]
\setlength{\epsfysize}{02in}
\centerline{\epsffile{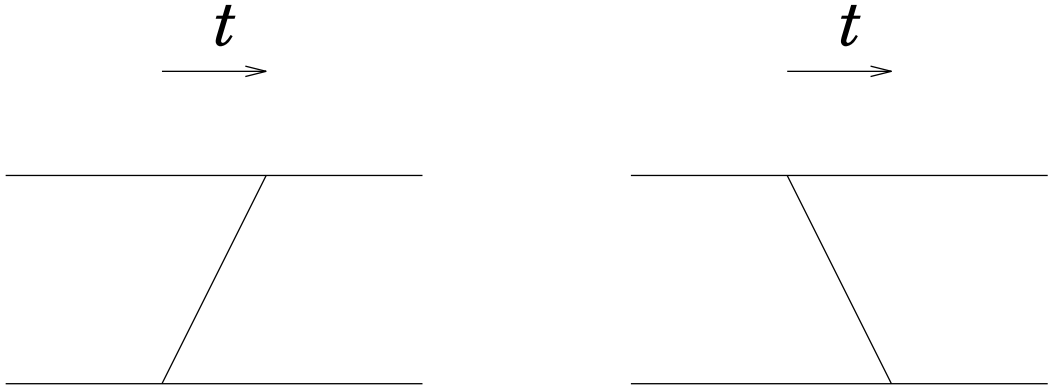}}
\caption[]{
The lowest diagrams for a scattering amplitude in the $t$-ordered OFPT.
}
\label{fig:0,ET_OFPT}
\end{figure}

\begin{figure}[ht]
\setlength{\epsfysize}{02in}
\centerline{\epsffile{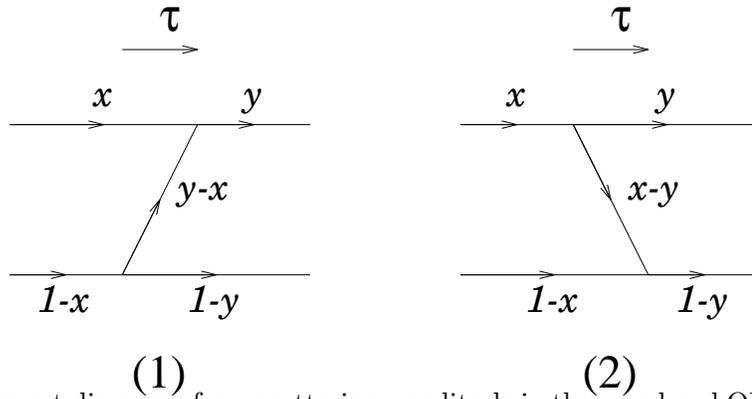}}
\caption[]{
The lowest diagrams for a scattering amplitude in the $\tau$-ordered OFPT.
Only lightcone plus(+) momentum fraction is shown.
}
\label{fig:0,LT_OFPT}
\end{figure}

\newpage

\begin{figure}[ht]
\setlength{\epsfysize}{01in}
\centerline{\epsffile{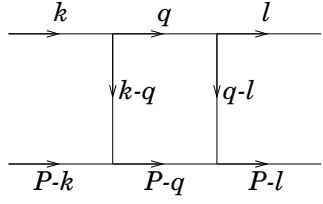}}
\caption[]{
The next-to-leading order ladder diagram for two-body scattering amplitude in CVPT.
}
\label{fig:1,CVPT}
\end{figure}

\begin{figure}[ht]
\setlength{\epsfysize}{06in}
\centerline{\epsffile{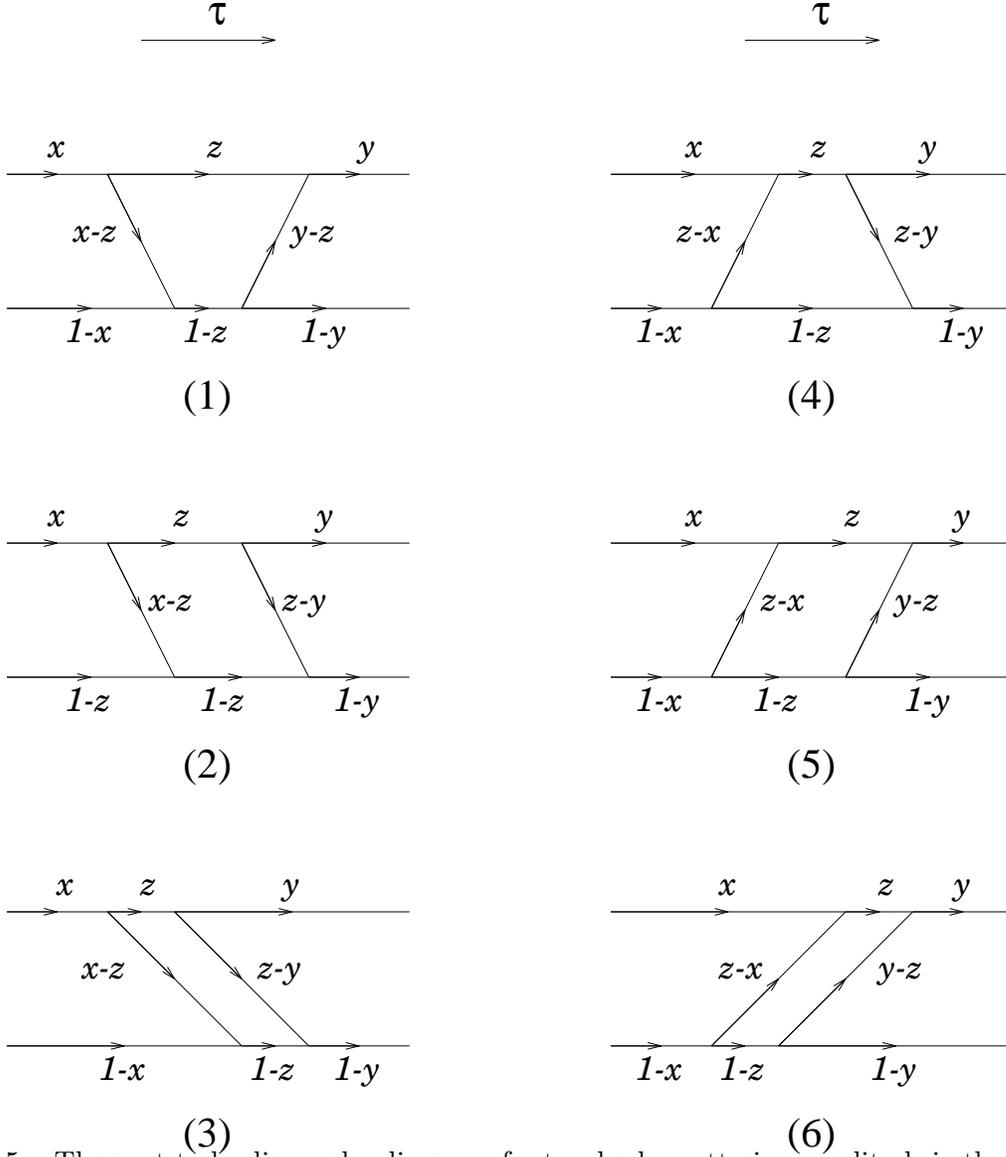}}
\caption[]{
The next-to-leading order diagrams for two-body scattering amplitude
in the $\tau$-ordered OFPT.
Only lightcone plus(+) momentum fraction is shown.
}
\label{fig:1,LT_OFPT}
\end{figure}

\newpage

\begin{figure}[ht]
\setlength{\epsfysize}{03in}
\centerline{\epsffile{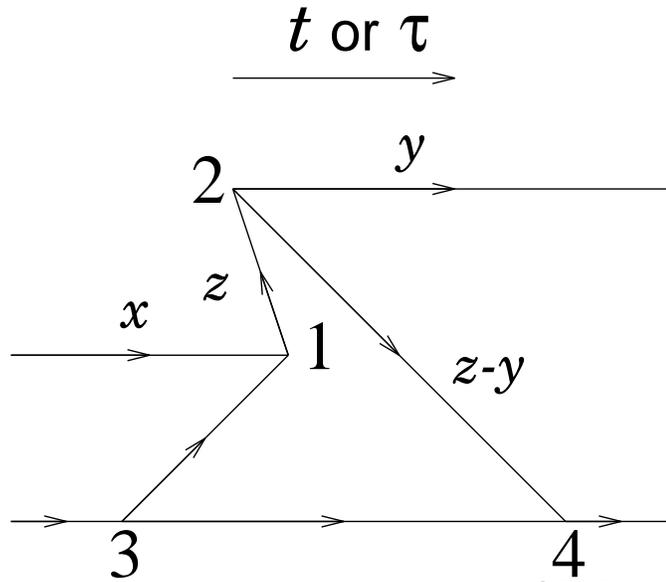}}
\caption[]{
A sample diagram which appears in the $t$-ordered OFPT but
does not appear in the $\tau$-ordered OFPT.
Only lightcone plus(+) momentum fraction is shown.
}
\label{fig:1,NULL}
\end{figure}

\newpage

\newpage

\newpage

\newpage

\newpage

\newpage

\newpage

\newpage

\newpage

\end{document}